\documentclass[showpacs,amsmath,
reprint,
aip]{revtex4-1}

\usepackage{graphicx}

\begin{document}

\title
{Optical control of individual carbon nanotube light emitters by spectral double resonance in silicon microdisk resonators}
\affiliation{Institute of Engineering Innovation, 
The University of Tokyo, Tokyo 113-8656, Japan}
\author{S.~Imamura}
\author{R.~Watahiki}
\author{R.~Miura}
\author{T.~Shimada}
\author{Y.~K.~Kato}
\email[Author to whom correspence should be addressed. Electronic mail: ]{ykato@sogo.t.u-tokyo.ac.jp}

\begin{abstract}
Single-walled carbon nanotubes have advantages as a nanoscale light source compatible with silicon photonics because they show room-temperature luminescence at telecom-wavelengths and can be directly synthesized on silicon substrates. Here we demonstrate integration of individual light-emitting carbon nanotubes with silicon microdisk resonators. Photons emitted from nanotubes are efficiently coupled to whispering gallery modes, circulating within the disks and lighting up their perimeters. Furthermore, we control such emission by tuning the excitation wavelength in and out of resonance with higher order modes in the same disk. Our results open up the possibilities of using nanotube emitters embedded in photonic circuits that are individually addressable through spectral double resonance. 
\end{abstract}

\maketitle

Advances in silicon photonics have enabled on-chip integration of various devices, \cite{Liang:2010, Reed:2010, Michel:2010, Leuthold:2010} expanding the capabilities of monolithic photonic circuits. For further scaling and increased functionality, however, integration of nanoscale emitters is desirable. Unfortunately, partly due to the fact that silicon is an indirect gap semiconductor, coupling of individually addressable nanoscale emitters to silicon photonics devices has remained a challenge.

In this regard, carbon nanotubes (CNTs) are promising because they are room-temperature telecom-band emitters \cite{O'Connell:2002, Weisman:2003} that can be directly synthesized on silicon \cite{Kong:1998} and be electrically driven.\cite{Misewich:2003, Chen:2005, Mann:2007, Xia:2008, Mueller:2010} In particular, as-grown air-suspended CNTs show excellent optical properties,\cite{Lefebvre:2003, Chen:2005, Mann:2007} making them ideal for use as individual emitters compared to on-substrate \cite{Xia:2008, Mueller:2010} or solution based materials.\cite{Watahiki:2012, Gaufres:2012}

As a silicon photonics device for coupling with air-suspended CNTs, microdisk resonators \cite{McCall:1992} have many desirable properties. They are a commonly used microcavity structure, supporting whispering gallery modes (WGMs) where optical waves are guided along the circumference of the disk by continuous internal reflection. Because they can have ultra-high quality factors \cite{Borselli:2005, Soltani:2007} and small mode volumes, WGMs are suited for coupling with nanoscale emitters such as quantum dots.\cite{Michler:2000,  Peter:2005, Mintairov:2008} It is relatively easy to tune the cavity resonance to the emitter wavelength by fabricating disks with an appropriate diameter, and the three-dimensional structure of post-supported microdisks \cite{McCall:1992, Michler:2000, Peter:2005} are compatible with air-suspended CNTs. Furthermore, they can be coupled to waveguides with high efficiency because they are traveling-wave resonators,\cite{Soltani:2007} making them a preferred choice compared to planar cavities \cite{Xia:2008} for integration in silicon photonics. 

Here we report on the integration of individual CNT emitters with silicon microdisks using fabrication processes compatible with standard silicon photonics. When photons emitted from nanotubes are coupled to the WGMs of the microdisks, emission enhancement occurs in a narrow spectral windows corresponding to the modes. Spatial imaging of the emission at the WGM wavelength shows the peripheral of the disk illuminated by the coupled CNT emission that circulates within the microdisk. In addition, we show that the excitation laser can also be coupled to whispering gallery modes in the same disk, allowing for non-local control over nanotube emission. 

\begin{figure}
\includegraphics{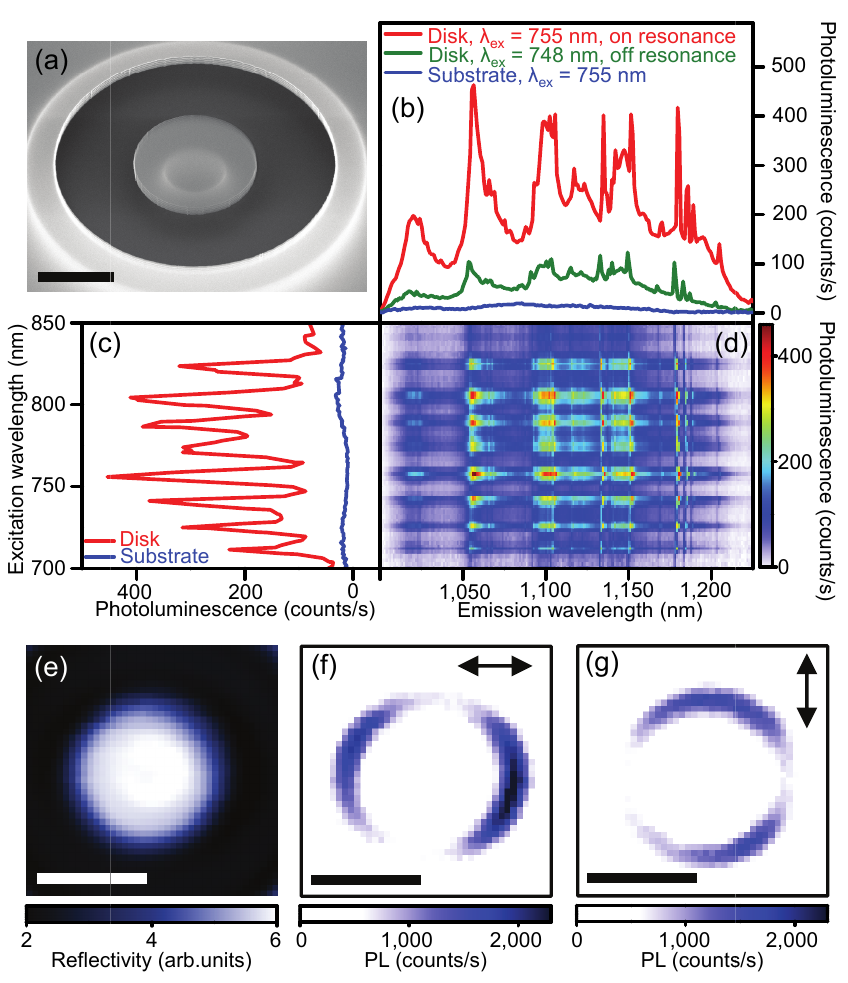}
\caption{\label{fig1}
(a) Scanning electron micrograph of an as-fabricated silicon microdisk. 
(b) PL spectra of a microdisk with a diameter of 2.93~$\mu$m under resonant excitation (red, $\lambda_\text{ex} = 755$~nm) and off-resonant excitation (green, $\lambda_\text{ex} = 748$~nm). Blue curve is a PL spectrum of the unpatterned substrate ($\lambda_\text{ex} = 755$~nm).
(c) PLE spectra of the microdisk (red) and unpatterned substrate (blue), obtained by integrating emission spectra over a spectral window of  1~nm centered around  1055~nm.
(d) PLE map. In (b)-(d), PL counts are normalized to $P=1$~mW to account for laser power changes with wavelength.
(e) Reflectivity image of the microdisk measured in  (b)-(g).
(f) and (g) PL images of the microdisk with horizontally and vertically polarized excitation, respectively. Images are obtained by integrating PL over a spectral window of  5~nm centered around  1055~nm. The polarization directions are schematically shown at the top-right corners of the images. The scale bars in (e)-(g) are 2~$\mu$m. 
}\end{figure}

We fabricate microdisks from a silicon-on-insulator wafer with a 260-nm-thick top Si layer and a 2000-nm-thick buried oxide layer. Electron beam lithography and dry-etching processes are used to form doughnut-shaped trenches in the top Si layer. In order to create a post-supported disk structure, the buried oxide layer is etched in buffered hydrofluoric acid. A scanning electron microscope image of a typical microdisk prior to nanotube growth is shown in Fig.~\ref{fig1}(a). 

 The devices are characterized with a home-built laser-scanning confocal photoluminescence (PL) microscopy system.\cite{Moritsubo:2010, Yasukochi:2011,Watahiki:2012} An output of a wavelength-tunable continuous-wave Ti:sapphire laser is focused onto the sample by an objective lens to a spot size of $\sim$1 $\mu$m. PL is coupled to a 300-mm spectrometer through a pinhole in a confocal configuration. An InGaAs photodiode array is used for detection and PL images are collected with a steering mirror. All measurements are done in air at room temperature.

 A PL spectrum of a microdisk taken with an excitation wavelength $\lambda_\text{ex} = 755$~nm and a power $P=1$~mW is shown as a red curve in Fig.~\ref{fig1}(b). A forest of resonances consisting of WGMs as well as Fabry-Perot modes \cite{Mintairov:2008} are visible within the silicon bandgap emission. 

An interesting observation is that the emission intensity decreases by a factor of $\sim$5 when $\lambda_\text{ex}$ is tuned to 748~nm [green curve in Fig.~\ref{fig1}(b)]. To investigate this strong excitation wavelength dependence, PL excitation (PLE) spectroscopy is performed [Fig.~\ref{fig1}(c)-(d)]. We interpret the cross hatch pattern in the PLE map [Fig.~\ref{fig1}(d)] as coupling of cavity modes at both excitation and emission wavelengths. The vertical streak-like features correspond to microdisk modes coupled to silicon emission, and the horizontal streaks represent modes coupled to the excitation laser. The intersections of vertical and horizontal streaks indicate simultaneous resonance of the excitation and emission, which allows control over PL intensity by about an order of magnitude [Fig.~\ref{fig1}(b)]. 

The PLE spectrum [Fig.~\ref{fig1}(c)] shows clear resonances with quality factors of $\sim$100, but precise mode assignment is difficult at these wavelengths because of high mode orders and strong absorption. Nevertheless, examination of reflectivity and PL images shows that PL is enhanced at the edges of the disk when the polarization matches the radial direction [Fig.~\ref{fig1}(e)-(g)]. Such a polarization dependence is consistent with what one expects for WGMs with transverse-electric polarization. The coupling of excitation laser to WGMs implies that the laser photons can propagate within the disk at resonance, allowing for non-local excitation.

In order to couple single CNTs to microdisk WGMs, it is necessary to choose a disk with an appropriate diameter such that a suitable WGM exists at the emission wavelength of nanotubes. In addition, as CNTs interact primarily with light polarized along their axis,\cite{Hartschuh:2003, Lefebvre:2004, Moritsubo:2010} it is also important to consider the mode polarization. WGMs can have either transverse-electric (TE) or transverse-magnetic (TM) polarization, and  we expect TE modes to have better coupling with the nanotubes since CNTs suspended onto the microdisks will be parallel to the substrate.

\begin{figure}
\includegraphics{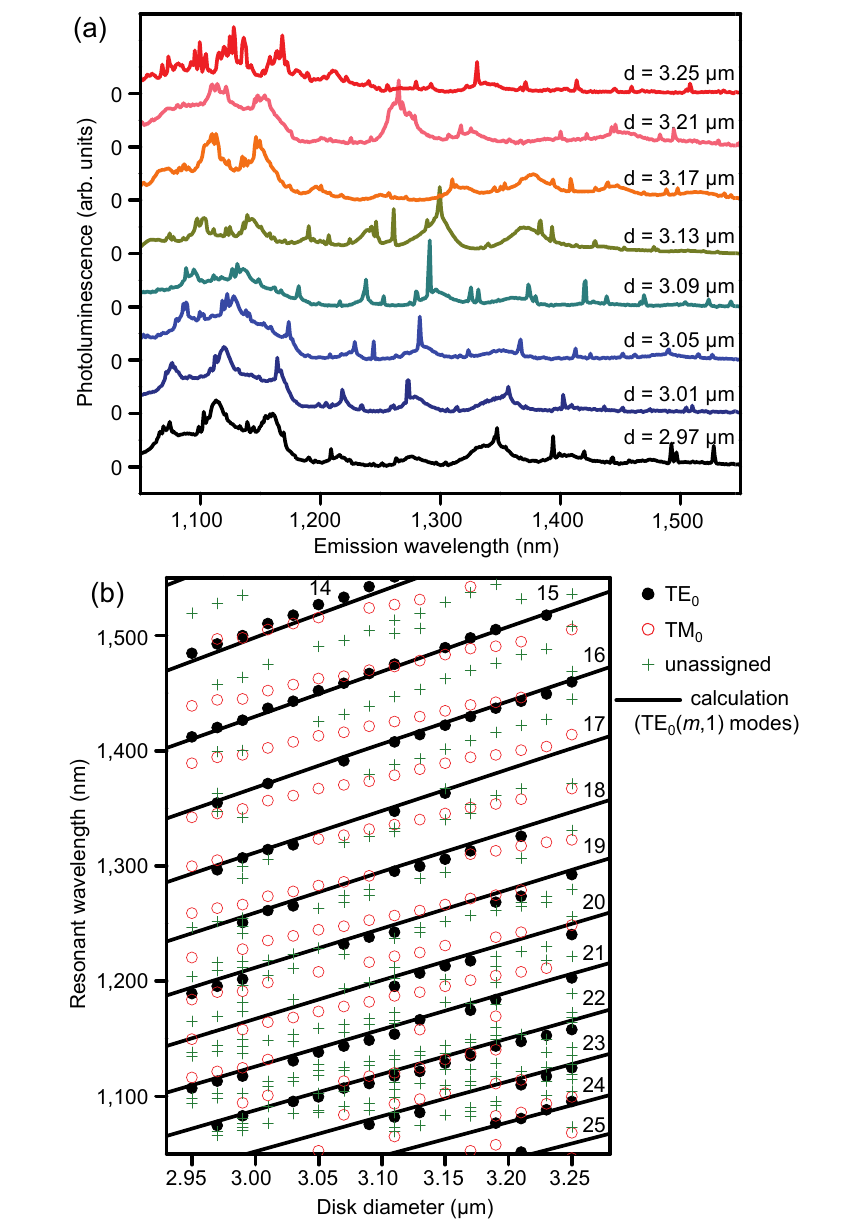}
\caption{\label{fig2}
(a) PL spectra of microdisks with diameters from 2.97~$\mu$m to 3.25~$\mu$m.
(b) Assignments of microdisk resonator modes. Lines show calculated wavelengths of fundamental TE modes using an approximated analytic model with no adjustable parameters. Numbers above the lines indicate mode index. Experimental data are obtained from spectra including those shown in (a). Filled circles are resonances identified as fundamental TE modes. Those assigned to fundamental TM modes are marked by open circles. Unassigned modes, which include higher order modes, are indicated by crosses.
}\end{figure}

We identify WGMs at CNT emission wavelengths by spin coating nanotube solution on the microdisks and performing PL measurements in a manner similar to experiments done on photonic crystal microcavities.\cite{Watahiki:2012} Figure~\ref{fig2}(a) shows spectra from a few microdisks with different diameters, and it can be seen that the mode wavelengths shift with diameter as expected. 

The disk diameter dependence of WGMs is summarized in Fig.~\ref{fig2}(b). We assign the TE modes by comparing the data to an analytical model with no adjustable parameters.\cite{Borselli:2005} The TM modes are identified by  a weaker dependence on the diameter, which is a result of their smaller effective index of refraction.
Within the wavelength range of interest  for coupling with CNTs ($1300$--$1500$~nm), we confirm the presence of a few TE modes. 

\begin{figure}
\includegraphics{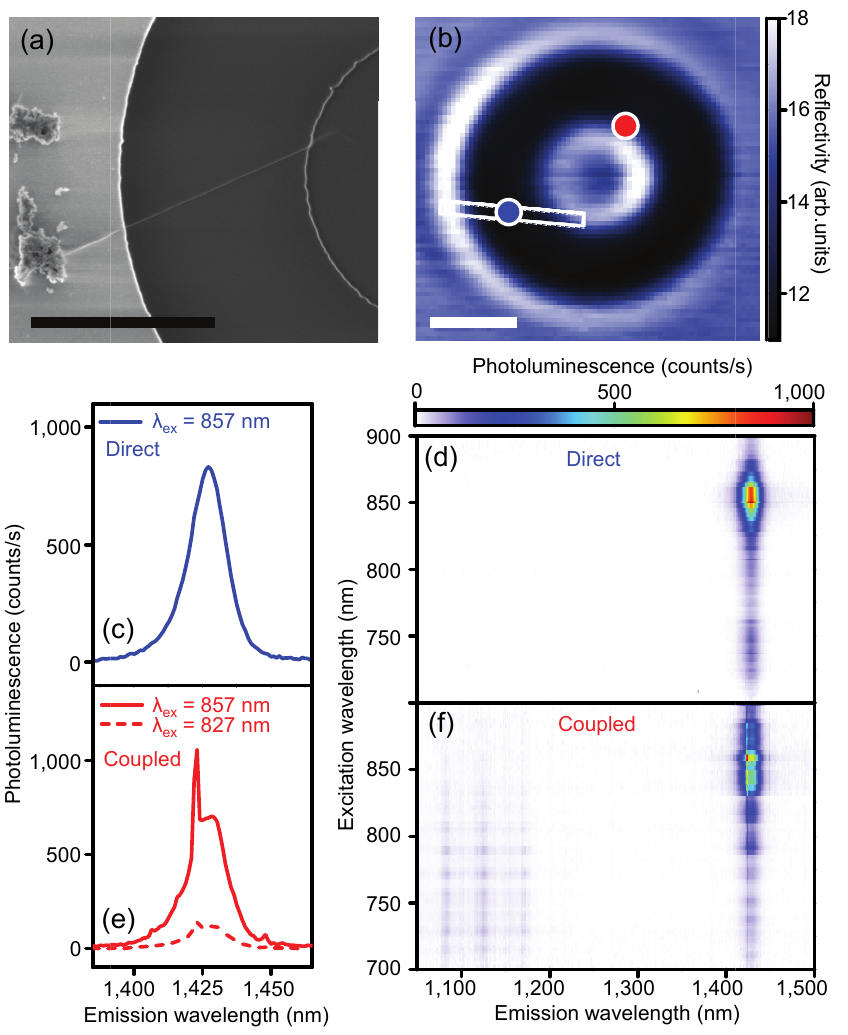}
\caption{\label{fig3}
(a) Scanning electron micrograph of a suspended nanotube attached to a microdisk.
(b) Reflectivity image of a device measured in (c)-(f). The disk diameter is 3.00~$\mu$m. $\lambda_\text{ex} = 800$~nm and $P=0.2$~mW are used. Blue and red dots indicate the positions at which the data in (c),(d) and (e),(f) are taken, respectively. The white box shows the position of the suspended nanotube. The scale bars in (a) and (b) are 2~$\mu$m. 
(c) Directly measured PL spectrum of the suspended nanotube for $\lambda_\text{ex} = 857$~nm.
(d) PLE map obtained by direct measurement of the suspended nanotube.
(e) PL spectrum of nanotube emission coupled to the microdisk for $\lambda_\text{ex} = 857$~nm (red solid line) and 827~nm (red broken line).
(f) PLE map of nanotube emission coupled to the microdisk.
In (c),(d) and (e),(f), the PL counts are normalized to laser powers of 10~$\mu$W and 500~$\mu$W, respectively. In (c),(d), laser is linearly polarized parallel to the nanotube, while in (e),(f), it is polarized along the radial direction of the disk.
}\end{figure}

We now investigate CNT emitters integrated with silicon microdisks [Fig.~\ref{fig3}(a)]. Another electron beam lithography step is performed to define catalyst windows at outer edges of the doughnut-shaped trenches. Single-walled CNTs are grown by chemical vapor deposition using ethanol as a carbon source.\cite{Maruyama:2002} Catalyst solution is prepared by ultrasonicating 5.0~mg of cobalt(II) acetate tetrahydrate and 50.0~mg of fumed silica in 40.0~g of ethanol. We spin coat and lift off the catalyst solution, and the samples are annealed in air for 5~minutes at 400$^\circ$C prior to growth. The samples are then placed in a quartz tube furnace, and the temperature is elevated to 800$^\circ$C while flowing Ar with 3\% H$_2$. Ethanol is introduced by bubbling the carrier gas for 10~minutes. With some probability, nanotubes can get suspended across the trench onto the microdisk. By controlling the size of the catalyst areas, we are able to obtain a reasonable yield of $\sim$10\% for microdisks with single suspended tubes. We note that the yield can be easily increased by placing more catalyst, if devices with multiple nanotubes are acceptable.

 Figure \ref{fig3}(b) shows a reflectivity image of a device with a CNT attached. Although the nanotube does not show up in this image, the white box indicates the position at where the nanotube is suspended. PL spectrum taken with the laser spot on the suspended tube is shown in Fig.~\ref{fig3}(c). The chiral index of this tube is determined to be $(10,8)$ by comparing the excitation map [Fig.~\ref{fig3}(d)] to tabulated data.\cite{Ohno:2006prb}

Surprisingly, the nanotube emission can also be observed at the opposite side of the disk [Fig.~\ref{fig3}(e)], although it is spatially separated from the nanotube by $\sim$3~$\mu$m. Notable is the sharp peak that appears on top of the nanotube spectrum. We identify this peak to be the 15th-order WGM with TE polarization [Fig.~\ref{fig2}(b)].  As the mode wavelength corresponds to an energy far below the bandgap of Si, we attribute the emergence of this peak to nanotube emission coupled to a WGM. 

The PLE map taken at the same position [Fig.~\ref{fig3}(f)] resembles that of the directly measured nanotube [Fig.~\ref{fig3}(d)], and in particular, the narrow WGM peak and the broader CNT emission peak are both maximized at the same excitation wavelength. This implies that the absorption characteristics are of the same origin for both WGM and CNT peaks, further supporting that the emission from the suspended nanotube is coupled to the WGM.  We also note that faint Si emission is visible at wavelengths shorter than 1200~nm in Fig.~\ref{fig3}(f). The contrast with bright  CNT emission demonstrates the advantage of using direct-gap nanotube emitters.

Different from the direct measurement of the suspended CNT, the PLE map for coupled emission shows strong modulation of intensity with changes in the excitation wavelength. As mentioned previously, there exist WGMs at these wavelengths as well, and they can cause enhancements of the excitation efficiency when excitation is resonant with a WGM. When both the excitation and emission are in resonance with WGMs, we meet the doubly resonant condition of the cavity. Similar to the case of Si PL, the emission is enhanced by a factor of $\sim$10 at double resonance compared to off-resonant conditions  [Fig.~\ref{fig3}(e)], allowing non-local control over nanotube emission through spectral tuning. 

\begin{figure}
\includegraphics{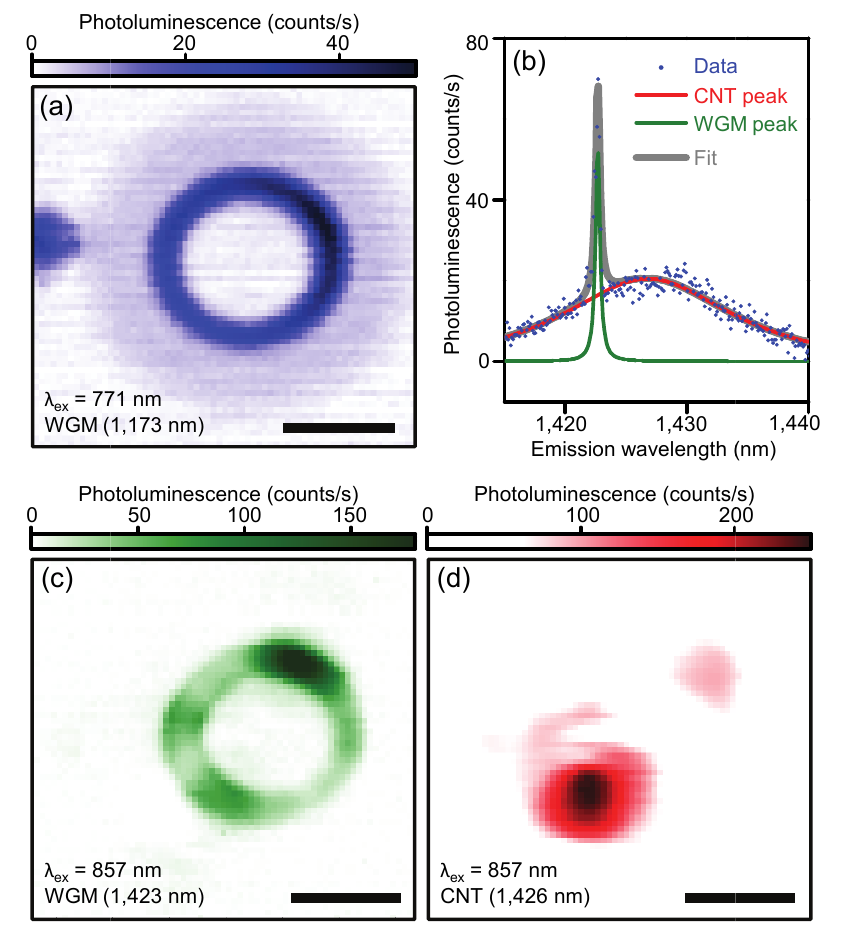}
\caption{\label{fig4}
(a) A PL image of Si emission coupled to a WGM, taken with $\lambda_\text{ex} = 771$~nm and $P=1.5$~mW. To construct this image, the spectra have been integrated from 1171~nm to 1175~nm.
(b) High resolution PL spectrum of nanotube emission coupled to a WGM. Blue dots are data and lines are Lorentzian fits as explained in the text. The peak values obtained from the fits are plotted in (c) and (d).
(c) An image of CNT emission coupled to a WGM.
(d) An image of direct CNT emission. For these images, $\lambda_\text{ex} = 857$~nm and $P=0.3$~mW with circular polarization are used. The scale bars in the images are 2~$\mu$m. 
}\end{figure}

The photons that couple to the WGM circulate within the microdisk, and this can be visualized through spectrally resolved PL imaging as in the case of WGM at Si emission wavelengths [Fig.~\ref{fig4}(a)]. Using a higher resolution grating, we collect PL spectra as we raster the laser spot over the scan area, and fit them with two Lorentzian functions corresponding to the WGM peak and the direct CNT emission peak [Fig.~\ref{fig4}(b)]. We find that the measured quality factor of the WGM is limited by spectrometer resolution to be $\sim$3000, which is already two orders of magnitude larger compared to a simple planar cavity.\cite{Xia:2008} In Fig.~\ref{fig4}(c) and \ref{fig4}(d), spectrally resolved PL  images corresponding to the WGM and direct CNT emission are presented. It is clear that nanotube emission coupled to the WGM circulates within the disk and illuminates the circumference [Fig.~\ref{fig4}(c)], showing that CNT emission is efficiently guided into the silicon photonic structure. 

The direct emission from the CNT is centered on the nanotube itself as expected [Fig.~\ref{fig4}(d)], although one can observe a lower intensity spot at the opposite side of the disk. We interpret the existence of such a remote spot in terms of in-plane Fabry-Perot mode within the microdisk.\cite{Mintairov:2008} Such an interpretation is also consistent with the image of the WGM [Fig.~\ref{fig4}(c)], which shows larger intensity near that spot.

Our results demonstrate the feasibility of integrating telecom-wavelength nanotube emitters in silicon photonics, opening up further possibilities for scaling down monolithic photonic circuits.  By fabricating arrays of microdisks with different diameters and thus different resonant wavelengths, it should be possible to address each microdisk by tuning the excitation wavelength. Furthermore, the WGMs that couple to emission will also differ, which would be useful for implementing on-chip wavelength-division multiplexing. Integration of single photon emitting CNTs \cite{Hogele:2008, Walden-Newman:2012} with silicon photonics may open up new opportunities for quantum optics on a chip.\cite{Politi:2008}

\begin{acknowledgments}
The authors acknowledge support from SCOPE, KAKENHI (21684016, 23104704, 24340066, 24654084), Asahi Glass Foundation,  KDDI Foundation, and the Photon Frontier Network Program of MEXT, Japan. The devices were fabricated at the Center for Nano Lithography \& Analysis at The University of Tokyo.
\end{acknowledgments}

\bibliography{Microdisk}

\begin{thebibliography}{30}%
\makeatletter
\providecommand \@ifxundefined [1]{%
 \@ifx{#1\undefined}
}%
\providecommand \@ifnum [1]{%
 \ifnum #1\expandafter \@firstoftwo
 \else \expandafter \@secondoftwo
 \fi
}%
\providecommand \@ifx [1]{%
 \ifx #1\expandafter \@firstoftwo
 \else \expandafter \@secondoftwo
 \fi
}%
\providecommand \natexlab [1]{#1}%
\providecommand \enquote  [1]{``#1''}%
\providecommand \bibnamefont  [1]{#1}%
\providecommand \bibfnamefont [1]{#1}%
\providecommand \citenamefont [1]{#1}%
\providecommand \href@noop [0]{\@secondoftwo}%
\providecommand \href [0]{\begingroup \@sanitize@url \@href}%
\providecommand \@href[1]{\@@startlink{#1}\@@href}%
\providecommand \@@href[1]{\endgroup#1\@@endlink}%
\providecommand \@sanitize@url [0]{\catcode `\\12\catcode `\$12\catcode
  `\&12\catcode `\#12\catcode `\^12\catcode `\_12\catcode `\%12\relax}%
\providecommand \@@startlink[1]{}%
\providecommand \@@endlink[0]{}%
\providecommand \url  [0]{\begingroup\@sanitize@url \@url }%
\providecommand \@url [1]{\endgroup\@href {#1}{\urlprefix }}%
\providecommand \urlprefix  [0]{URL }%
\providecommand \Eprint [0]{\href }%
\providecommand \doibase [0]{http://dx.doi.org/}%
\providecommand \selectlanguage [0]{\@gobble}%
\providecommand \bibinfo  [0]{\@secondoftwo}%
\providecommand \bibfield  [0]{\@secondoftwo}%
\providecommand \translation [1]{[#1]}%
\providecommand \BibitemOpen [0]{}%
\providecommand \bibitemStop [0]{}%
\providecommand \bibitemNoStop [0]{.\EOS\space}%
\providecommand \EOS [0]{\spacefactor3000\relax}%
\providecommand \BibitemShut  [1]{\csname bibitem#1\endcsname}%
\let\auto@bib@innerbib\@empty
\bibitem [{\citenamefont {Liang}\ and\ \citenamefont
  {Bowers}(2010)}]{Liang:2010}%
  \BibitemOpen
  \bibfield  {author} {\bibinfo {author} {\bibfnamefont {D.}~\bibnamefont
  {Liang}}\ and\ \bibinfo {author} {\bibfnamefont {J.~E.}\ \bibnamefont
  {Bowers}},\ }\href {\doibase 10.1038/nphoton.2010.167} {\bibfield  {journal}
  {\bibinfo  {journal} {Nature Photon.}\ }\textbf {\bibinfo {volume} {4}},\
  \bibinfo {pages} {511} (\bibinfo {year} {2010})}\BibitemShut {NoStop}%
\bibitem [{\citenamefont {Reed}\ \emph {et~al.}(2010)\citenamefont {Reed},
  \citenamefont {Mashanovich}, \citenamefont {Gardes},\ and\ \citenamefont
  {Thomson}}]{Reed:2010}%
  \BibitemOpen
  \bibfield  {author} {\bibinfo {author} {\bibfnamefont {G.~T.}\ \bibnamefont
  {Reed}}, \bibinfo {author} {\bibfnamefont {G.}~\bibnamefont {Mashanovich}},
  \bibinfo {author} {\bibfnamefont {F.~Y.}\ \bibnamefont {Gardes}}, \ and\
  \bibinfo {author} {\bibfnamefont {D.~J.}\ \bibnamefont {Thomson}},\ }\href
  {\doibase 10.1038/nphoton.2010.179} {\bibfield  {journal} {\bibinfo
  {journal} {Nature Photon.}\ }\textbf {\bibinfo {volume} {4}},\ \bibinfo
  {pages} {518} (\bibinfo {year} {2010})}\BibitemShut {NoStop}%
\bibitem [{\citenamefont {Michel}, \citenamefont {Liu},\ and\ \citenamefont
  {Kimerling}(2010)}]{Michel:2010}%
  \BibitemOpen
  \bibfield  {author} {\bibinfo {author} {\bibfnamefont {J.}~\bibnamefont
  {Michel}}, \bibinfo {author} {\bibfnamefont {J.}~\bibnamefont {Liu}}, \ and\
  \bibinfo {author} {\bibfnamefont {L.~C.}\ \bibnamefont {Kimerling}},\ }\href
  {\doibase 10.1038/nphoton.2010.157} {\bibfield  {journal} {\bibinfo
  {journal} {Nature Photon.}\ }\textbf {\bibinfo {volume} {4}},\ \bibinfo
  {pages} {527} (\bibinfo {year} {2010})}\BibitemShut {NoStop}%
\bibitem [{\citenamefont {Leuthold}, \citenamefont {Koos},\ and\ \citenamefont
  {Freude}(2010)}]{Leuthold:2010}%
  \BibitemOpen
  \bibfield  {author} {\bibinfo {author} {\bibfnamefont {J.}~\bibnamefont
  {Leuthold}}, \bibinfo {author} {\bibfnamefont {C.}~\bibnamefont {Koos}}, \
  and\ \bibinfo {author} {\bibfnamefont {W.}~\bibnamefont {Freude}},\ }\href
  {\doibase 10.1038/nphoton.2010.185} {\bibfield  {journal} {\bibinfo
  {journal} {Nature Photon.}\ }\textbf {\bibinfo {volume} {4}},\ \bibinfo
  {pages} {535} (\bibinfo {year} {2010})}\BibitemShut {NoStop}%
\bibitem [{\citenamefont {O'Connell}\ \emph {et~al.}(2002)\citenamefont
  {O'Connell}, \citenamefont {Bachilo}, \citenamefont {Huffman}, \citenamefont
  {Moore}, \citenamefont {Strano}, \citenamefont {Haroz}, \citenamefont
  {Rialon}, \citenamefont {Boul}, \citenamefont {Noon}, \citenamefont
  {Kittrell}, \citenamefont {Ma}, \citenamefont {Hauge}, \citenamefont
  {Weisman},\ and\ \citenamefont {Smalley}}]{O'Connell:2002}%
  \BibitemOpen
  \bibfield  {author} {\bibinfo {author} {\bibfnamefont {M.~J.}\ \bibnamefont
  {O'Connell}}, \bibinfo {author} {\bibfnamefont {S.~M.}\ \bibnamefont
  {Bachilo}}, \bibinfo {author} {\bibfnamefont {C.~B.}\ \bibnamefont
  {Huffman}}, \bibinfo {author} {\bibfnamefont {V.~C.}\ \bibnamefont {Moore}},
  \bibinfo {author} {\bibfnamefont {M.~S.}\ \bibnamefont {Strano}}, \bibinfo
  {author} {\bibfnamefont {E.~H.}\ \bibnamefont {Haroz}}, \bibinfo {author}
  {\bibfnamefont {K.~L.}\ \bibnamefont {Rialon}}, \bibinfo {author}
  {\bibfnamefont {P.~J.}\ \bibnamefont {Boul}}, \bibinfo {author}
  {\bibfnamefont {W.~H.}\ \bibnamefont {Noon}}, \bibinfo {author}
  {\bibfnamefont {C.}~\bibnamefont {Kittrell}}, \bibinfo {author}
  {\bibfnamefont {J.}~\bibnamefont {Ma}}, \bibinfo {author} {\bibfnamefont
  {R.~H.}\ \bibnamefont {Hauge}}, \bibinfo {author} {\bibfnamefont {R.~B.}\
  \bibnamefont {Weisman}}, \ and\ \bibinfo {author} {\bibfnamefont {R.~E.}\
  \bibnamefont {Smalley}},\ }\href {\doibase 10.1126/science.1072631}
  {\bibfield  {journal} {\bibinfo  {journal} {Science}\ }\textbf {\bibinfo
  {volume} {297}},\ \bibinfo {pages} {593} (\bibinfo {year}
  {2002})}\BibitemShut {NoStop}%
\bibitem [{\citenamefont {Weisman}\ and\ \citenamefont
  {Bachilo}(2003)}]{Weisman:2003}%
  \BibitemOpen
  \bibfield  {author} {\bibinfo {author} {\bibfnamefont {R.~B.}\ \bibnamefont
  {Weisman}}\ and\ \bibinfo {author} {\bibfnamefont {S.~M.}\ \bibnamefont
  {Bachilo}},\ }\href {\doibase 10.1021/nl034428i} {\bibfield  {journal}
  {\bibinfo  {journal} {Nano Lett.}\ }\textbf {\bibinfo {volume} {3}},\
  \bibinfo {pages} {1235} (\bibinfo {year} {2003})}\BibitemShut {NoStop}%
\bibitem [{\citenamefont {Kong}\ \emph {et~al.}(1998)\citenamefont {Kong},
  \citenamefont {Soh}, \citenamefont {Cassell}, \citenamefont {Quate},\ and\
  \citenamefont {Dai}}]{Kong:1998}%
  \BibitemOpen
  \bibfield  {author} {\bibinfo {author} {\bibfnamefont {J.}~\bibnamefont
  {Kong}}, \bibinfo {author} {\bibfnamefont {H.~T.}\ \bibnamefont {Soh}},
  \bibinfo {author} {\bibfnamefont {A.~M.}\ \bibnamefont {Cassell}}, \bibinfo
  {author} {\bibfnamefont {C.~F.}\ \bibnamefont {Quate}}, \ and\ \bibinfo
  {author} {\bibfnamefont {H.}~\bibnamefont {Dai}},\ }\href {\doibase
  10.1038/27632} {\bibfield  {journal} {\bibinfo  {journal} {Nature}\ }\textbf
  {\bibinfo {volume} {395}},\ \bibinfo {pages} {878} (\bibinfo {year}
  {1998})}\BibitemShut {NoStop}%
\bibitem [{\citenamefont {Misewich}\ \emph {et~al.}(2003)\citenamefont
  {Misewich}, \citenamefont {Martel}, \citenamefont {Avouris}, \citenamefont
  {Tsang}, \citenamefont {Heinze},\ and\ \citenamefont
  {Tersoff}}]{Misewich:2003}%
  \BibitemOpen
  \bibfield  {author} {\bibinfo {author} {\bibfnamefont {J.~A.}\ \bibnamefont
  {Misewich}}, \bibinfo {author} {\bibfnamefont {R.}~\bibnamefont {Martel}},
  \bibinfo {author} {\bibfnamefont {P.}~\bibnamefont {Avouris}}, \bibinfo
  {author} {\bibfnamefont {J.~C.}\ \bibnamefont {Tsang}}, \bibinfo {author}
  {\bibfnamefont {S.}~\bibnamefont {Heinze}}, \ and\ \bibinfo {author}
  {\bibfnamefont {J.}~\bibnamefont {Tersoff}},\ }\href {\doibase
  10.1126/science.1081294} {\bibfield  {journal} {\bibinfo  {journal}
  {Science}\ }\textbf {\bibinfo {volume} {300}},\ \bibinfo {pages} {783}
  (\bibinfo {year} {2003})}\BibitemShut {NoStop}%
\bibitem [{\citenamefont {Chen}\ \emph {et~al.}(2005)\citenamefont {Chen},
  \citenamefont {Perebeinos}, \citenamefont {Freitag}, \citenamefont {Tsang},
  \citenamefont {Fu}, \citenamefont {Liu},\ and\ \citenamefont
  {Avouris}}]{Chen:2005}%
  \BibitemOpen
  \bibfield  {author} {\bibinfo {author} {\bibfnamefont {J.}~\bibnamefont
  {Chen}}, \bibinfo {author} {\bibfnamefont {V.}~\bibnamefont {Perebeinos}},
  \bibinfo {author} {\bibfnamefont {M.}~\bibnamefont {Freitag}}, \bibinfo
  {author} {\bibfnamefont {J.}~\bibnamefont {Tsang}}, \bibinfo {author}
  {\bibfnamefont {Q.}~\bibnamefont {Fu}}, \bibinfo {author} {\bibfnamefont
  {J.}~\bibnamefont {Liu}}, \ and\ \bibinfo {author} {\bibfnamefont
  {P.}~\bibnamefont {Avouris}},\ }\href {\doibase 10.1126/science.1119177}
  {\bibfield  {journal} {\bibinfo  {journal} {Science}\ }\textbf {\bibinfo
  {volume} {310}},\ \bibinfo {pages} {1171} (\bibinfo {year}
  {2005})}\BibitemShut {NoStop}%
\bibitem [{\citenamefont {Mann}\ \emph {et~al.}(2007)\citenamefont {Mann},
  \citenamefont {Kato}, \citenamefont {Kinkhabwala}, \citenamefont {Pop},
  \citenamefont {Cao}, \citenamefont {Wang}, \citenamefont {Zhang},
  \citenamefont {Wang}, \citenamefont {Guo},\ and\ \citenamefont
  {Dai}}]{Mann:2007}%
  \BibitemOpen
  \bibfield  {author} {\bibinfo {author} {\bibfnamefont {D.}~\bibnamefont
  {Mann}}, \bibinfo {author} {\bibfnamefont {Y.~K.}\ \bibnamefont {Kato}},
  \bibinfo {author} {\bibfnamefont {A.}~\bibnamefont {Kinkhabwala}}, \bibinfo
  {author} {\bibfnamefont {E.}~\bibnamefont {Pop}}, \bibinfo {author}
  {\bibfnamefont {J.}~\bibnamefont {Cao}}, \bibinfo {author} {\bibfnamefont
  {X.}~\bibnamefont {Wang}}, \bibinfo {author} {\bibfnamefont {L.}~\bibnamefont
  {Zhang}}, \bibinfo {author} {\bibfnamefont {Q.}~\bibnamefont {Wang}},
  \bibinfo {author} {\bibfnamefont {J.}~\bibnamefont {Guo}}, \ and\ \bibinfo
  {author} {\bibfnamefont {H.}~\bibnamefont {Dai}},\ }\href {\doibase
  10.1038/nnano.2006.169} {\bibfield  {journal} {\bibinfo  {journal} {Nature
  Nanotech.}\ }\textbf {\bibinfo {volume} {2}},\ \bibinfo {pages} {33}
  (\bibinfo {year} {2007})}\BibitemShut {NoStop}%
\bibitem [{\citenamefont {Xia}\ \emph {et~al.}(2008)\citenamefont {Xia},
  \citenamefont {Steiner}, \citenamefont {Lin},\ and\ \citenamefont
  {Avouris}}]{Xia:2008}%
  \BibitemOpen
  \bibfield  {author} {\bibinfo {author} {\bibfnamefont {F.}~\bibnamefont
  {Xia}}, \bibinfo {author} {\bibfnamefont {M.}~\bibnamefont {Steiner}},
  \bibinfo {author} {\bibfnamefont {Y.-M.}\ \bibnamefont {Lin}}, \ and\
  \bibinfo {author} {\bibfnamefont {P.}~\bibnamefont {Avouris}},\ }\href
  {\doibase 10.1038/nnano.2008.241} {\bibfield  {journal} {\bibinfo  {journal}
  {Nature Nanotech.}\ }\textbf {\bibinfo {volume} {3}},\ \bibinfo {pages} {609}
  (\bibinfo {year} {2008})}\BibitemShut {NoStop}%
\bibitem [{\citenamefont {Mueller}\ \emph {et~al.}(2010)\citenamefont
  {Mueller}, \citenamefont {Kinoshita}, \citenamefont {Steiner}, \citenamefont
  {Perebeinos}, \citenamefont {Bol}, \citenamefont {Farmer},\ and\
  \citenamefont {Avouris}}]{Mueller:2010}%
  \BibitemOpen
  \bibfield  {author} {\bibinfo {author} {\bibfnamefont {T.}~\bibnamefont
  {Mueller}}, \bibinfo {author} {\bibfnamefont {M.}~\bibnamefont {Kinoshita}},
  \bibinfo {author} {\bibfnamefont {M.}~\bibnamefont {Steiner}}, \bibinfo
  {author} {\bibfnamefont {V.}~\bibnamefont {Perebeinos}}, \bibinfo {author}
  {\bibfnamefont {A.~A.}\ \bibnamefont {Bol}}, \bibinfo {author} {\bibfnamefont
  {D.~B.}\ \bibnamefont {Farmer}}, \ and\ \bibinfo {author} {\bibfnamefont
  {P.}~\bibnamefont {Avouris}},\ }\href {\doibase 10.1038/nnano.2009.319}
  {\bibfield  {journal} {\bibinfo  {journal} {Nature Nanotech.}\ }\textbf
  {\bibinfo {volume} {5}},\ \bibinfo {pages} {27} (\bibinfo {year}
  {2010})}\BibitemShut {NoStop}%
\bibitem [{\citenamefont {Lefebvre}, \citenamefont {Homma},\ and\ \citenamefont
  {Finnie}(2003)}]{Lefebvre:2003}%
  \BibitemOpen
  \bibfield  {author} {\bibinfo {author} {\bibfnamefont {J.}~\bibnamefont
  {Lefebvre}}, \bibinfo {author} {\bibfnamefont {Y.}~\bibnamefont {Homma}}, \
  and\ \bibinfo {author} {\bibfnamefont {P.}~\bibnamefont {Finnie}},\ }\href
  {\doibase 10.1103/PhysRevLett.90.217401} {\bibfield  {journal} {\bibinfo
  {journal} {Phys. Rev. Lett.}\ }\textbf {\bibinfo {volume} {90}},\ \bibinfo
  {pages} {217401} (\bibinfo {year} {2003})}\BibitemShut {NoStop}%
\bibitem [{\citenamefont {Watahiki}\ \emph {et~al.}(2012)\citenamefont
  {Watahiki}, \citenamefont {Shimada}, \citenamefont {Zhao}, \citenamefont
  {Chiashi}, \citenamefont {Iwamoto}, \citenamefont {Arakawa}, \citenamefont
  {Maruyama},\ and\ \citenamefont {Kato}}]{Watahiki:2012}%
  \BibitemOpen
  \bibfield  {author} {\bibinfo {author} {\bibfnamefont {R.}~\bibnamefont
  {Watahiki}}, \bibinfo {author} {\bibfnamefont {T.}~\bibnamefont {Shimada}},
  \bibinfo {author} {\bibfnamefont {P.}~\bibnamefont {Zhao}}, \bibinfo {author}
  {\bibfnamefont {S.}~\bibnamefont {Chiashi}}, \bibinfo {author} {\bibfnamefont
  {S.}~\bibnamefont {Iwamoto}}, \bibinfo {author} {\bibfnamefont
  {Y.}~\bibnamefont {Arakawa}}, \bibinfo {author} {\bibfnamefont
  {S.}~\bibnamefont {Maruyama}}, \ and\ \bibinfo {author} {\bibfnamefont
  {Y.~K.}\ \bibnamefont {Kato}},\ }\href {\doibase 10.1063/1.4757876}
  {\bibfield  {journal} {\bibinfo  {journal} {Appl. Phys. Lett.}\ }\textbf
  {\bibinfo {volume} {101}},\ \bibinfo {eid} {141124} (\bibinfo {year}
  {2012})}\BibitemShut {NoStop}%
\bibitem [{\citenamefont {Gaufr\`{e}s}\ \emph {et~al.}(2012)\citenamefont
  {Gaufr\`{e}s}, \citenamefont {Izard}, \citenamefont {Noury}, \citenamefont
  {Le~Roux}, \citenamefont {Rasigade}, \citenamefont {Beck},\ and\
  \citenamefont {Vivien}}]{Gaufres:2012}%
  \BibitemOpen
  \bibfield  {author} {\bibinfo {author} {\bibfnamefont {E.}~\bibnamefont
  {Gaufr\`{e}s}}, \bibinfo {author} {\bibfnamefont {N.}~\bibnamefont {Izard}},
  \bibinfo {author} {\bibfnamefont {A.}~\bibnamefont {Noury}}, \bibinfo
  {author} {\bibfnamefont {X.}~\bibnamefont {Le~Roux}}, \bibinfo {author}
  {\bibfnamefont {G.}~\bibnamefont {Rasigade}}, \bibinfo {author}
  {\bibfnamefont {A.}~\bibnamefont {Beck}}, \ and\ \bibinfo {author}
  {\bibfnamefont {L.}~\bibnamefont {Vivien}},\ }\href {\doibase
  10.1021/nn204924n} {\bibfield  {journal} {\bibinfo  {journal} {ACS Nano}\
  }\textbf {\bibinfo {volume} {6}},\ \bibinfo {pages} {3813} (\bibinfo {year}
  {2012})}\BibitemShut {NoStop}%
\bibitem [{\citenamefont {McCall}\ \emph {et~al.}(1992)\citenamefont {McCall},
  \citenamefont {Levi}, \citenamefont {Slusher}, \citenamefont {Pearton},\ and\
  \citenamefont {Logan}}]{McCall:1992}%
  \BibitemOpen
  \bibfield  {author} {\bibinfo {author} {\bibfnamefont {S.~L.}\ \bibnamefont
  {McCall}}, \bibinfo {author} {\bibfnamefont {A.~F.~J.}\ \bibnamefont {Levi}},
  \bibinfo {author} {\bibfnamefont {R.~E.}\ \bibnamefont {Slusher}}, \bibinfo
  {author} {\bibfnamefont {S.~J.}\ \bibnamefont {Pearton}}, \ and\ \bibinfo
  {author} {\bibfnamefont {R.~A.}\ \bibnamefont {Logan}},\ }\href {\doibase
  10.1063/1.106688} {\bibfield  {journal} {\bibinfo  {journal} {Appl. Phys.
  Lett.}\ }\textbf {\bibinfo {volume} {60}},\ \bibinfo {pages} {289} (\bibinfo
  {year} {1992})}\BibitemShut {NoStop}%
\bibitem [{\citenamefont {Borselli}, \citenamefont {Johnson},\ and\
  \citenamefont {Painter}(2005)}]{Borselli:2005}%
  \BibitemOpen
  \bibfield  {author} {\bibinfo {author} {\bibfnamefont {M.}~\bibnamefont
  {Borselli}}, \bibinfo {author} {\bibfnamefont {T.}~\bibnamefont {Johnson}}, \
  and\ \bibinfo {author} {\bibfnamefont {O.}~\bibnamefont {Painter}},\ }\href
  {\doibase 10.1364/OPEX.13.001515} {\bibfield  {journal} {\bibinfo  {journal}
  {Opt. Express}\ }\textbf {\bibinfo {volume} {13}},\ \bibinfo {pages} {1515}
  (\bibinfo {year} {2005})}\BibitemShut {NoStop}%
\bibitem [{\citenamefont {Soltani}, \citenamefont {Yegnanarayanan},\ and\
  \citenamefont {Adibi}(2007)}]{Soltani:2007}%
  \BibitemOpen
  \bibfield  {author} {\bibinfo {author} {\bibfnamefont {M.}~\bibnamefont
  {Soltani}}, \bibinfo {author} {\bibfnamefont {S.}~\bibnamefont
  {Yegnanarayanan}}, \ and\ \bibinfo {author} {\bibfnamefont {A.}~\bibnamefont
  {Adibi}},\ }\href {\doibase 10.1364/OE.15.004694} {\bibfield  {journal}
  {\bibinfo  {journal} {Opt. Express}\ }\textbf {\bibinfo {volume} {15}},\
  \bibinfo {pages} {4694} (\bibinfo {year} {2007})}\BibitemShut {NoStop}%
\bibitem [{\citenamefont {Michler}\ \emph {et~al.}(2000)\citenamefont
  {Michler}, \citenamefont {Kiraz}, \citenamefont {Becher}, \citenamefont
  {Schoenfeld}, \citenamefont {Petroff}, \citenamefont {Zhang}, \citenamefont
  {Hu},\ and\ \citenamefont {Imamo\u{g}lu}}]{Michler:2000}%
  \BibitemOpen
  \bibfield  {author} {\bibinfo {author} {\bibfnamefont {P.}~\bibnamefont
  {Michler}}, \bibinfo {author} {\bibfnamefont {A.}~\bibnamefont {Kiraz}},
  \bibinfo {author} {\bibfnamefont {C.}~\bibnamefont {Becher}}, \bibinfo
  {author} {\bibfnamefont {W.~V.}\ \bibnamefont {Schoenfeld}}, \bibinfo
  {author} {\bibfnamefont {P.~M.}\ \bibnamefont {Petroff}}, \bibinfo {author}
  {\bibfnamefont {L.}~\bibnamefont {Zhang}}, \bibinfo {author} {\bibfnamefont
  {E.}~\bibnamefont {Hu}}, \ and\ \bibinfo {author} {\bibfnamefont
  {A.}~\bibnamefont {Imamo\u{g}lu}},\ }\href {\doibase
  10.1126/science.290.5500.2282} {\bibfield  {journal} {\bibinfo  {journal}
  {Science}\ }\textbf {\bibinfo {volume} {290}},\ \bibinfo {pages} {2282}
  (\bibinfo {year} {2000})}\BibitemShut {NoStop}%
\bibitem [{\citenamefont {Peter}\ \emph {et~al.}(2005)\citenamefont {Peter},
  \citenamefont {Senellart}, \citenamefont {Martrou}, \citenamefont
  {Lema\^itre}, \citenamefont {Hours}, \citenamefont {G\'erard},\ and\
  \citenamefont {Bloch}}]{Peter:2005}%
  \BibitemOpen
  \bibfield  {author} {\bibinfo {author} {\bibfnamefont {E.}~\bibnamefont
  {Peter}}, \bibinfo {author} {\bibfnamefont {P.}~\bibnamefont {Senellart}},
  \bibinfo {author} {\bibfnamefont {D.}~\bibnamefont {Martrou}}, \bibinfo
  {author} {\bibfnamefont {A.}~\bibnamefont {Lema\^itre}}, \bibinfo {author}
  {\bibfnamefont {J.}~\bibnamefont {Hours}}, \bibinfo {author} {\bibfnamefont
  {J.~M.}\ \bibnamefont {G\'erard}}, \ and\ \bibinfo {author} {\bibfnamefont
  {J.}~\bibnamefont {Bloch}},\ }\href {\doibase 10.1103/PhysRevLett.95.067401}
  {\bibfield  {journal} {\bibinfo  {journal} {Phys. Rev. Lett.}\ }\textbf
  {\bibinfo {volume} {95}},\ \bibinfo {pages} {067401} (\bibinfo {year}
  {2005})}\BibitemShut {NoStop}%
\bibitem [{\citenamefont {Mintairov}\ \emph {et~al.}(2008)\citenamefont
  {Mintairov}, \citenamefont {Chu}, \citenamefont {He}, \citenamefont
  {Blokhin}, \citenamefont {Nadtochy}, \citenamefont {Maximov}, \citenamefont
  {Tokranov}, \citenamefont {Oktyabrsky},\ and\ \citenamefont
  {Merz}}]{Mintairov:2008}%
  \BibitemOpen
  \bibfield  {author} {\bibinfo {author} {\bibfnamefont {A.~M.}\ \bibnamefont
  {Mintairov}}, \bibinfo {author} {\bibfnamefont {Y.}~\bibnamefont {Chu}},
  \bibinfo {author} {\bibfnamefont {Y.}~\bibnamefont {He}}, \bibinfo {author}
  {\bibfnamefont {S.}~\bibnamefont {Blokhin}}, \bibinfo {author} {\bibfnamefont
  {A.}~\bibnamefont {Nadtochy}}, \bibinfo {author} {\bibfnamefont
  {M.}~\bibnamefont {Maximov}}, \bibinfo {author} {\bibfnamefont
  {V.}~\bibnamefont {Tokranov}}, \bibinfo {author} {\bibfnamefont
  {S.}~\bibnamefont {Oktyabrsky}}, \ and\ \bibinfo {author} {\bibfnamefont
  {J.~L.}\ \bibnamefont {Merz}},\ }\href {\doibase 10.1103/PhysRevB.77.195322}
  {\bibfield  {journal} {\bibinfo  {journal} {Phys. Rev. B}\ }\textbf {\bibinfo
  {volume} {77}},\ \bibinfo {pages} {195322} (\bibinfo {year}
  {2008})}\BibitemShut {NoStop}%
\bibitem [{\citenamefont {Moritsubo}\ \emph {et~al.}(2010)\citenamefont
  {Moritsubo}, \citenamefont {Murai}, \citenamefont {Shimada}, \citenamefont
  {Murakami}, \citenamefont {Chiashi}, \citenamefont {Maruyama},\ and\
  \citenamefont {Kato}}]{Moritsubo:2010}%
  \BibitemOpen
  \bibfield  {author} {\bibinfo {author} {\bibfnamefont {S.}~\bibnamefont
  {Moritsubo}}, \bibinfo {author} {\bibfnamefont {T.}~\bibnamefont {Murai}},
  \bibinfo {author} {\bibfnamefont {T.}~\bibnamefont {Shimada}}, \bibinfo
  {author} {\bibfnamefont {Y.}~\bibnamefont {Murakami}}, \bibinfo {author}
  {\bibfnamefont {S.}~\bibnamefont {Chiashi}}, \bibinfo {author} {\bibfnamefont
  {S.}~\bibnamefont {Maruyama}}, \ and\ \bibinfo {author} {\bibfnamefont
  {Y.~K.}\ \bibnamefont {Kato}},\ }\href {\doibase
  10.1103/PhysRevLett.104.247402} {\bibfield  {journal} {\bibinfo  {journal}
  {Phys. Rev. Lett.}\ }\textbf {\bibinfo {volume} {104}},\ \bibinfo {pages}
  {247402} (\bibinfo {year} {2010})}\BibitemShut {NoStop}%
\bibitem [{\citenamefont {Yasukochi}\ \emph {et~al.}(2011)\citenamefont
  {Yasukochi}, \citenamefont {Murai}, \citenamefont {Moritsubo}, \citenamefont
  {Shimada}, \citenamefont {Chiashi}, \citenamefont {Maruyama},\ and\
  \citenamefont {Kato}}]{Yasukochi:2011}%
  \BibitemOpen
  \bibfield  {author} {\bibinfo {author} {\bibfnamefont {S.}~\bibnamefont
  {Yasukochi}}, \bibinfo {author} {\bibfnamefont {T.}~\bibnamefont {Murai}},
  \bibinfo {author} {\bibfnamefont {S.}~\bibnamefont {Moritsubo}}, \bibinfo
  {author} {\bibfnamefont {T.}~\bibnamefont {Shimada}}, \bibinfo {author}
  {\bibfnamefont {S.}~\bibnamefont {Chiashi}}, \bibinfo {author} {\bibfnamefont
  {S.}~\bibnamefont {Maruyama}}, \ and\ \bibinfo {author} {\bibfnamefont
  {Y.~K.}\ \bibnamefont {Kato}},\ }\href {\doibase 10.1103/PhysRevB.84.121409}
  {\bibfield  {journal} {\bibinfo  {journal} {Phys. Rev. B}\ }\textbf {\bibinfo
  {volume} {84}},\ \bibinfo {pages} {121409(R)} (\bibinfo {year}
  {2011})}\BibitemShut {NoStop}%
\bibitem [{\citenamefont {Hartschuh}\ \emph {et~al.}(2003)\citenamefont
  {Hartschuh}, \citenamefont {Pedrosa}, \citenamefont {Novotny},\ and\
  \citenamefont {Krauss}}]{Hartschuh:2003}%
  \BibitemOpen
  \bibfield  {author} {\bibinfo {author} {\bibfnamefont {A.}~\bibnamefont
  {Hartschuh}}, \bibinfo {author} {\bibfnamefont {H.~N.}\ \bibnamefont
  {Pedrosa}}, \bibinfo {author} {\bibfnamefont {L.}~\bibnamefont {Novotny}}, \
  and\ \bibinfo {author} {\bibfnamefont {T.~D.}\ \bibnamefont {Krauss}},\
  }\href {\doibase 10.1126/science.1087118} {\bibfield  {journal} {\bibinfo
  {journal} {Science}\ }\textbf {\bibinfo {volume} {301}},\ \bibinfo {pages}
  {1354} (\bibinfo {year} {2003})}\BibitemShut {NoStop}%
\bibitem [{\citenamefont {Lefebvre}\ \emph {et~al.}(2004)\citenamefont
  {Lefebvre}, \citenamefont {Fraser}, \citenamefont {Finnie},\ and\
  \citenamefont {Homma}}]{Lefebvre:2004}%
  \BibitemOpen
  \bibfield  {author} {\bibinfo {author} {\bibfnamefont {J.}~\bibnamefont
  {Lefebvre}}, \bibinfo {author} {\bibfnamefont {J.~M.}\ \bibnamefont
  {Fraser}}, \bibinfo {author} {\bibfnamefont {P.}~\bibnamefont {Finnie}}, \
  and\ \bibinfo {author} {\bibfnamefont {Y.}~\bibnamefont {Homma}},\ }\href
  {\doibase 10.1103/PhysRevB.69.075403} {\bibfield  {journal} {\bibinfo
  {journal} {Phys. Rev. B}\ }\textbf {\bibinfo {volume} {69}},\ \bibinfo
  {pages} {075403} (\bibinfo {year} {2004})}\BibitemShut {NoStop}%
\bibitem [{\citenamefont {Maruyama}\ \emph {et~al.}(2002)\citenamefont
  {Maruyama}, \citenamefont {Kojima}, \citenamefont {Miyauchi}, \citenamefont
  {Chiashi},\ and\ \citenamefont {Kohno}}]{Maruyama:2002}%
  \BibitemOpen
  \bibfield  {author} {\bibinfo {author} {\bibfnamefont {S.}~\bibnamefont
  {Maruyama}}, \bibinfo {author} {\bibfnamefont {R.}~\bibnamefont {Kojima}},
  \bibinfo {author} {\bibfnamefont {Y.}~\bibnamefont {Miyauchi}}, \bibinfo
  {author} {\bibfnamefont {S.}~\bibnamefont {Chiashi}}, \ and\ \bibinfo
  {author} {\bibfnamefont {M.}~\bibnamefont {Kohno}},\ }\href {\doibase
  10.1016/S0009-2614(02)00838-2} {\bibfield  {journal} {\bibinfo  {journal}
  {Chem. Phys. Lett.}\ }\textbf {\bibinfo {volume} {360}},\ \bibinfo {pages}
  {229} (\bibinfo {year} {2002})}\BibitemShut {NoStop}%
\bibitem [{\citenamefont {Ohno}\ \emph {et~al.}(2006)\citenamefont {Ohno},
  \citenamefont {Iwasaki}, \citenamefont {Murakami}, \citenamefont {Kishimoto},
  \citenamefont {Maruyama},\ and\ \citenamefont {Mizutani}}]{Ohno:2006prb}%
  \BibitemOpen
  \bibfield  {author} {\bibinfo {author} {\bibfnamefont {Y.}~\bibnamefont
  {Ohno}}, \bibinfo {author} {\bibfnamefont {S.}~\bibnamefont {Iwasaki}},
  \bibinfo {author} {\bibfnamefont {Y.}~\bibnamefont {Murakami}}, \bibinfo
  {author} {\bibfnamefont {S.}~\bibnamefont {Kishimoto}}, \bibinfo {author}
  {\bibfnamefont {S.}~\bibnamefont {Maruyama}}, \ and\ \bibinfo {author}
  {\bibfnamefont {T.}~\bibnamefont {Mizutani}},\ }\href {\doibase
  10.1103/PhysRevB.73.235427} {\bibfield  {journal} {\bibinfo  {journal} {Phys.
  Rev. B}\ }\textbf {\bibinfo {volume} {73}},\ \bibinfo {pages} {235427}
  (\bibinfo {year} {2006})}\BibitemShut {NoStop}%
\bibitem [{\citenamefont {H\"ogele}\ \emph {et~al.}(2008)\citenamefont
  {H\"ogele}, \citenamefont {Galland}, \citenamefont {Winger},\ and\
  \citenamefont {Imamo\u{g}lu}}]{Hogele:2008}%
  \BibitemOpen
  \bibfield  {author} {\bibinfo {author} {\bibfnamefont {A.}~\bibnamefont
  {H\"ogele}}, \bibinfo {author} {\bibfnamefont {C.}~\bibnamefont {Galland}},
  \bibinfo {author} {\bibfnamefont {M.}~\bibnamefont {Winger}}, \ and\ \bibinfo
  {author} {\bibfnamefont {A.}~\bibnamefont {Imamo\u{g}lu}},\ }\href {\doibase
  10.1103/PhysRevLett.100.217401} {\bibfield  {journal} {\bibinfo  {journal}
  {Phys. Rev. Lett.}\ }\textbf {\bibinfo {volume} {100}},\ \bibinfo {pages}
  {217401} (\bibinfo {year} {2008})}\BibitemShut {NoStop}%
\bibitem [{\citenamefont {Walden-Newman}, \citenamefont {Sarpkaya},\ and\
  \citenamefont {Strauf}(2012)}]{Walden-Newman:2012}%
  \BibitemOpen
  \bibfield  {author} {\bibinfo {author} {\bibfnamefont {W.}~\bibnamefont
  {Walden-Newman}}, \bibinfo {author} {\bibfnamefont {I.}~\bibnamefont
  {Sarpkaya}}, \ and\ \bibinfo {author} {\bibfnamefont {S.}~\bibnamefont
  {Strauf}},\ }\href {\doibase 10.1021/nl204402v} {\bibfield  {journal}
  {\bibinfo  {journal} {Nano Lett.}\ }\textbf {\bibinfo {volume} {12}},\
  \bibinfo {pages} {1934} (\bibinfo {year} {2012})}\BibitemShut {NoStop}%
\bibitem [{\citenamefont {Politi}\ \emph {et~al.}(2008)\citenamefont {Politi},
  \citenamefont {Cryan}, \citenamefont {Rarity}, \citenamefont {Yu},\ and\
  \citenamefont {O'Brien}}]{Politi:2008}%
  \BibitemOpen
  \bibfield  {author} {\bibinfo {author} {\bibfnamefont {A.}~\bibnamefont
  {Politi}}, \bibinfo {author} {\bibfnamefont {M.~J.}\ \bibnamefont {Cryan}},
  \bibinfo {author} {\bibfnamefont {J.~G.}\ \bibnamefont {Rarity}}, \bibinfo
  {author} {\bibfnamefont {S.}~\bibnamefont {Yu}}, \ and\ \bibinfo {author}
  {\bibfnamefont {J.~L.}\ \bibnamefont {O'Brien}},\ }\href {\doibase
  10.1126/science.1155441} {\bibfield  {journal} {\bibinfo  {journal}
  {Science}\ }\textbf {\bibinfo {volume} {320}},\ \bibinfo {pages} {646}
  (\bibinfo {year} {2008})}\BibitemShut {NoStop}%
\end{thebibliography}%

\end{document}